\documentclass[usenatbib]{emulateapj}
\usepackage{graphicx}
\usepackage{amssymb,amsmath} 

\usepackage{color}

\newcommand{\msun}{M_{\odot}}

\begin{document}
\title{Constraining the dark energy equation of state using {\it LISA} observations of spinning Massive Black Hole binaries.}
\author{Antoine Petiteau\altaffilmark{1}\email{}, Stanislav Babak\altaffilmark{1}\email{} \& Alberto Sesana\altaffilmark{1}\email{}}

\altaffiltext{1}{Max-Planck-Institut fuer Gravitationsphysik,
Albert-Einstein-Institut, Am Muhlenberg 1, D-14476 Golm bei Potsdam, Germany}


\begin{abstract} 
Gravitational wave signals from coalescing Massive Black Hole (MBH) binaries could be used 
as standard sirens to measure cosmological parameters.
The future space based gravitational wave observatory Laser Interferometer Space Antenna ({\it LISA}) 
will detect up to a hundred of those events, providing very accurate measurements of their  luminosity distances.
To constrain the cosmological parameters we also need to  measure the redshift of the galaxy 
(or cluster of galaxies) hosting the merger. This requires the identification of a distinctive electromagnetic 
event associated to the binary coalescence. However, putative electromagnetic signatures may be too weak 
to be observed. Instead, we study here the possibility of constraining the cosmological parameters by enforcing
statistical consistency between all the possible hosts detected within the measurement error box of a few 
dozen of low redshift ($z<3$) events. We construct MBH populations using merger tree realizations
of the dark matter hierarchy in a $\Lambda$CDM Universe, and we use data from the Millennium 
simulation to model the galaxy distribution in the {\it LISA} error box. We show that, assuming that 
all the other cosmological parameters are known, the parameter $w$ describing the dark energy equation 
of state can be constrained to a 4-8\% level (2$\sigma$ error), competitive with current
uncertainties obtained by type Ia supernovae measurements, providing an independent test of our 
cosmological model.

\end{abstract}

\vskip 0.5true cm


\maketitle

\section{Introduction}
\label{S:Intro}

The Laser Interferometer Space Antenna \citep[{{\it{LISA}}},][]{1997CQGra..14.1399D}
is a space based gravitational wave (GW) observatory which is expected to be launched in 2022+.
One of its central scientific goals is to provide information about the cosmic evolution
of massive black holes (MBHs). It is, infact,  now widely recognized that MBHs are fundamental
building blocks in the process of galaxy formation and evolution; 
they are ubiquitous in nearby galaxy nuclei  
\citep[see, e.g., ][]{mago98}, and their masses tightly correlate with
the properties of their host \citep[][and references therein]{gult09}. 
In popular $\Lambda$CDM cosmologies, structure formation proceeds
in a hierarchical fashion \citep{wr78}, through a sequence of merging events. 
If MBHs are common in galaxy centers at all epochs, as implied by the
notion that galaxies harbor active nuclei for a short period of
their lifetime \citep{hr93}, then a large number 
of MBH binaries are expected to form during cosmic history. 
{\it LISA} is expected to observe the GW driven inspiral and final coalescence of such 
MBH binaries out to very high redshift with high signal-to-noise ratio (SNR), 
allowing very accurate measurements of the 
binary parameters. The collective properties of the set of the observed coalescing binaries  
will carry invaluable information for astrophysics, making possible to constrain models of MBH
formation and growth \citep{plowman10, Gair:2010bx,  sesanaetal10}.


Besides astrophysical applications, coalescing MBHs could be used 
as standard sirens \citep{Schutz:1986gp, Holz:2005df, SpinBBHLangHugues2006, Arun:2007hu, Arun:2008xf, Lang:2008gh, VanDenBroeck:2010fp}.
The high strength of the GW signals allows us to measure the luminosity distance with a precision of less than 
a percent at redshift $z=1$ (neglecting weak lensing). However, we need an electromagnetic 
identification of the host in order to measure the source redshift and be able to do cosmography. 
If the event is nearby ($z < 0.4$), 
then we have a very good localization of the source on the sky and we can identify a single cluster of galaxies 
hosting the merger. As we go to higher redshifts, {\it LISA} sky localization abilities become quite poor: 
a typical sky resolution for an equal mass $10^6\msun$ inspiralling MBH  binary at $z=1$ is 20-30 arcminutes
a side at $2\sigma$  
\citep{Trias:2007fp, Lang:2008gh, Arun:2008zn}, which  is in general not sufficient to uniquely identify the host 
of the GW event. There is, therefore, a growing interest in identifying putative electromagnetic signatures associated 
to the MBH binary before and/or after the final GW driven coalescence \citep[for a review, see][and references therein]{Schnittman:2010wy}.
Electromagnetic anomalies observed before or after the coalescence within the {\it LISA} measurements error box may 
allow us to identify the host and to make a redshift measurement. However, most of the
proposed electromagnetic counterparts are rather weak (below the Eddington limit), and in case of dry
mergers (no cold gas efficiently funneled into the remnant nucleus) we do not expect any distinctive electromagnetic 
transient. This brings us back to the original idea by \cite{Schutz:1986gp} to consider each galaxy within the {\it LISA} 
measurement error box as a potential host candidate. The idea is that, by cross-correlating several GW events, only one 
galaxy (cluster of galaxies) in each error box will give us a consistent set of parameters describing the Universe. 
The effectiveness of this method has been demonstrated by \cite{MacLeod:2007jd} in the context of the Hubble 
constant determination by means of low redshift ($z<0.2$) extreme mass ratio inspirals.

We use the hierarchical MBH formation model suggested by \cite{Volonteri:2010py} to generate catalogs of coalescing
MBH binaries along the cosmic history. This model predicts $\sim 100$  MBHs mergers observable by {\it LISA} 
in three years, in the redshift range $[0:5]$. We do not use sources beyond redshift $z=3$ 
due to difficulties of measuring galaxy redshifts 
beyond that threshold\footnote{There are other reasons for not going beyond $z=3$ which we will discuss later.}.  
We model the galaxy distribution in the Universe using the Millennium simulation \citep{Springel:2005nw}. 
For each coalescing MBH in our catalog, we select a host galaxy in the Millennium run snapshot  
closest in redshift to the actual redshift of the event. 
For each galaxy in the snapshot, we compute the apparent magnitude in some observable band, and
we create a catalog of redshift measurements of all the observable potential host candidates.
Note that typical observed mergers involve $10^4-10^6\; M_{\odot}$ MBHs, which implies 
\citep[using the black hole mass--bulge relations, see, e.g., ][]{gult09}
relatively light galaxies. However, observed galaxies 
are heavy due to selection effects: roughly speaking, mass reflects luminosity, so that at high redshifts we can 
observe only very massive (luminous) galaxies. Therefore, the actual host might not be (and often is not) 
among the observed galaxies. The important fact is the self-similarity of the density distribution:
the local density distribution for all galaxies and the density distribution for heavy galaxies are 
quite similar, which allow us to infer the likelihood of the host redshift on the basis of redshift 
measurements of the luminous galaxies only.
 
We assume that the GW source parameter measurements (GW likelihoods) are represented by multivariate Gaussian distributions
around the true values, with the variance-covariance matrix defined by the inverse of the Fisher matrix. 
This is a good approximation in the case of Gaussian instrumental noise and large SNR.  
At $z \ge 0.25$ the uncertainty in the luminosity distance ($D_L$) is dominated by 
weak lensing due to the extended distribution of dark matter halos between us and the GW source. 
In this paper we combine the luminosity distance errors given by GW measurements 
and weak lensing, referring to them as GW+WL errors. We use two estimations of the weak lensing error
(i) from \cite{Shapiro:2009sr}  and (ii) from \cite{Wang:2002qc}. 

In order to evaluate the error box we need to assume some prior on the cosmological parameters.
In this exploratory study, we assume that we know all the cosmological parameters but the effective equation of state 
for the dark energy, described by the parameter $w$ (which could be the case by the time {\it LISA} will fly). 
In a follow up paper we will relax this assumption by including also the Hubble constant and the
matter and dark energy content of the Universe as free parameters. 
We take the prior range for $w$ from the seven-year WMAP analysis \citep{2010arXiv1001.4538K}. 
We show that using statistical methods $w$ can be constrained to a 4-8\% level (2$\sigma$ error
),
providing an effective method for estimating the dark energy equation of state.
We also show that this result  depends weakly on the prior range and could serve as an 
independent way of measuring the dark energy equation of state, with respect to canonical methods
employing observations of type Ia supernovae \citep{riess98}.

The paper is structured as follows.  In Section \ref{S:Sim} we spell out explicitly all the details of the
adopted cosmological model and of the Bayesian analytical framework. In Section \ref{S:astro}
we give more insights on the MBH population model and on the galaxy distributions extracted from the 
Millennium database. In Section~\ref{S:Obs} we describe our 
simulated GW and electromagnetic observations. We give results of our simulations under different 
assumptions about weak lensing, depth of the follow up 
electromagnetic surveys, etc. in Section~\ref{S:Res}. We summarize our findings in Section \ref{S:Sum}.

\section{Analytical framework}
\label{S:Sim}

\subsection{Cosmological description of the Universe}
\label{SS:Sim:Cosmo}

We assume the standard $\Lambda\textrm{CDM}$ model, 
which describes our Universe as the sum of two non-interacting components: 
(i) a pressureless component corresponding to all visible and dark matter,  
(ii) a dark energy component with current effective equation of state corresponding to the $\Lambda-$term $p = -\epsilon$. Current estimates based on SN1a observations and anisotropy measurements in the cosmic microwave background  
\citep{riess98,2010arXiv1001.4538K} tell us that about 70\% of the Universe energy content is in the form of the  dark energy. 
The evolution of the Universe is therefore described by the expansion equation 
 \begin{equation}
 H^2 = H_0^2 \left[ \Omega_m^0 (1+z)^3 + 
\Omega_{de}^0 \exp \left( 
3\int_0^z dz \frac{1+\omega(z)}{1+z}\right)  \right].
 \end{equation}
 where $H = \dot{a}/a$ ($a$ being  the lengthscale of the Universe) 
is the Hubble expansion parameter and $H_0$ is its current value ($t=0$),
 $\Omega_m$ and $\Omega_{de}$ are the ratios of the matter density and the dark energy 
density to the critical density, and $\omega(z)$ describes the effective dark energy equation of state 
as a function of $z$. We assume that the Universe is spatially flat,
the luminosity distance is therefore computed as 
 \begin{equation}
 D_L = (1+z) \int_0^z \frac{dz'}{H(z')}.
 \label{Eq:DLofz}
 \end{equation} 
 In our simulations we fix all parameters (assuming that they are known exactly) to the currently 
 estimated mean values: $H_0  = 73.0 \ \textrm{km}\times \textrm{s}^{-1}\times\textrm{Mpc}^{-1}$, $ \Omega_m = 0.25,\; \Omega_{de} = 0.75$.
 We also simplify the form of $\omega(z)$ for which we will assume $\omega = -1 - w$, where $w$
 is a constant~\footnote{Here we use notations for the dark energy equation of state adopted in the WMAP data analysis \citep{2010arXiv1001.4538K}.}. 
We choose the value  $w=0$ to simulate our Universe which is what has been used in the Millennium simulation (see below). 

\subsection{Methodology and working plan}
\label{SS:Sim:Cosmo}

Our aim is to show that we can constrain $w$  via GW observations of spinning MBH binaries, using a Bayesian
framework. Let us consider $j=1,..,N_{\rm ev}$ GW observations. For each event we can infer the probability of a parameter 
$w$, given the collected data $s$, using Bayes theorem:   
\begin{equation}
P_j(w | s) = \frac{p_0(w) P_j(s | w)}{E_j}. 
\label{Eq:Bayes}
\end{equation}
Here $P_j(w | s)$ is the posterior probability of the parameter $w$, $P_j(s | w)$ is the likelihood of the observation
$s$ given the parameter $w$, $p_0(w)$ in the prior knowledge of $w$ and $E_j$ is defined as 
\begin{equation}
E_j = \int p_0(w) P_j(s | w) dw.
\end{equation}
The likelihood $P_j(w | s)$ must be appropriately specialized to our problem. We want to exploit GW observations to
constrain $w$ through the distance - redshift ($D_L-z$) relation as given by (\ref{Eq:DLofz}).
\begin{itemize} 
\item The distance $D_L$ is provided by the GW observations: the GW 
signal carries information about the parameters of the binary, including its location on the sky and its luminosity 
distance. All those parameters can be extracted using latest data analysis methods 
\citep{2010PhRvD..81j4016P, Cornish:2006ms}. The measurements errors are encoded in the GW likelihood
\footnote{Through the paper, with GW likelihood we mean the likelihood of the {\it LISA} data to contain the GW signal 
with a given parameters, not to be confused with the likelihood $P_j(s | w)$ defined in the Bayes theorem.} function 
$\mathcal{L}(D_L, \theta, \phi, \vec{\lambda})$, where $\{\theta, \phi\}$  are the ecliptic coordinates 
of the source and $\vec{\lambda}$ represents all the other parameters characterizing MBH binary (spins and their 
orientation, masses, orientation of the orbit and MBHs position at the beginning of observations). 
When estimating $D_L$ weak lensing can not be neglected. In fact the error 
coming from the weak lensing (causing fluctuations in the brightness of the GW source which gives an uncertainty 
in the luminosity distance) dominates over the GW error starting from redshift $z \sim 0.25$ (see figure \ref{F:WL}).
\item The redshift measurement does not rely on any distinctive electromagnetic signature related to the GW event. 
We extract a redshift probability distribution of the host from the clustering properties of the galaxies 
falling withing the GW+WL error box. This defines an astrophysical prior  $p(\theta, \phi, z)$ for a given galaxy 
in the measurement error box to be the host of coalescing 
binary. To translate the measured $D_L$ and uncertainty $\Delta D_L$  of the GW event into a 
corresponding $z$ and $\Delta z$ for the candidate host galaxies in the sky we use the 
prior knowledge of $p_0(w)$ obtained from WMAP.
\end{itemize}

The likelihood in equation (\ref{Eq:Bayes}) can therefore be written as
\begin{eqnarray}
P_j(s | w) = \int \mathcal{L}_j\left[D_L(z,w), \theta, \phi,\vec{\lambda}\right] p(\vec{\lambda}) 
p_j(\theta, \phi, z)\; d\vec{\lambda}\; d\theta\; d\phi \; dz,\nonumber\\
{}
\label{Eq:like}
\end{eqnarray}
where we have introduced the priors $p(\vec{\lambda}) $ 
on the parameters $\vec{\lambda}$ (which we assume in this paper to be uniform). 
It is convenient to change the variable of integration from $D_L$ to $z$.  Since we have assumed 
uniform priors on $\vec{\lambda}$, we can marginalize the likelihood  
over those parameters~\footnote{Here this corresponds to the projection of the Fisher matrix to three dimensional
parameter space of sky location $\theta, \phi$ and luminosity distance $D_L$.} to obtain:
\begin{equation}
P_j(s|w) = \int \pi_j\left[D_L(z,w), \theta, \phi\right] p_j(\theta, \phi, z)\; d\theta\; d\phi\;  dz,
\label{Eq:poster}
\end{equation}
where we denoted the marginalized GW likelihood as $\pi_j\left[D_L(z,w), \theta, \phi\right]$. Practically, we limit 
the integration to the size of the error box (in principle the integration should be taken over the 
whole range of parameters but we found that considering the $2\sigma$ error box is sufficient). 

We assume that the error in luminosity distance from the weak lensing is not correlated with the GW measurements, 
hence the integral in equation (\ref{Eq:poster}) can be performed over the sky ($\{\theta, \phi\}$) first, and then 
over the redshift. 
We also found that the correlation between $D_L$ and the sky position coming from the GW observations is not 
important for events at $z<0.5$.  Plugging equation (\ref{Eq:poster}) into equation (\ref{Eq:Bayes}) defines the 
posterior distribution of $w$ for a single GW event (as indicated by the index $j$). Assuming that 
all $N_{\rm ev}$ GW events are independent, the combined posterior probability is
\begin{equation}
P(w) = \frac{p_0(w) \prod_{j=1}^{N_{\rm ev}} P_j(s| w)} {\int p_0(w) \prod_{j=1}^{N_{\rm ev}} P_j(s | w) dw}. 
\label{E:finalpost}
\end{equation}

To evaluate $w$ through equation (\ref{E:finalpost}) we therefore need:
\begin{itemize}
\item a MBH binary population model defining the properties of the $N_{\rm ev}$ coalescing systems;
\item the spatial distribution of galaxies within a volume comparable with the combined GW+WL measurement error box;
\item the measurement errors associated to GW observations of coalescing MBH binaries 
(defining $\mathcal{L}_j(D_L, \theta, \phi, \vec{\lambda})$);
\item an estimation of spectroscopic survey capabilities to construct the galaxy 
redshift distribution within the GW+WL measurement error box (defining $p_j(\theta, \phi, z)$).  
\end{itemize}
We will consider these points individually in the next two sections.

\section{Astrophysical background}
\label{S:astro}
\subsection{Massive black hole binary population}
\label{SS:Sim:MBHBpop}

To generate populations of MBH binaries in the Universe, we use the results of merger tree
simulations described in details in \cite{vhm03}. MBHs grow hierarchically,
starting from a distribution of seed black holes at high redshift, through a sequence of merger
and accretion episodes.
Two distinctive type of seeds have been proposed in the literature.
Light ($M\sim 100\msun$) seed are thought to be the remnant of Population III (POPIII) stars
\citep{Madau:2001sc}, whereas heavy seeds form following instabilities occurring in massive
protogalactic disks.
In the model proposed by Begelman Volonteri \& Rees \citep[][hereafter BVR model]{bvr06},  
a `quasistar' forms at the center of the protogalaxy, eventually collapsing into a seed BH that
efficiently accretes from the quasistar envelope, resulting in a final mass $M\sim$ few $\times 10^4\msun$. 
Here we use the model recently suggested by Volonteri \& Begelman \citep[][hereafter VB model]{Volonteri:2010py},
which combines the two above prescriptions by mixing light and heavy initial seeds. 
This model predicts $\sim30-50$ events per year in the
redshift range $0<z<3$, relevant to this study. 
\begin{figure}[!htbp]
\center
\includegraphics[width=0.45\textwidth, clip=true]{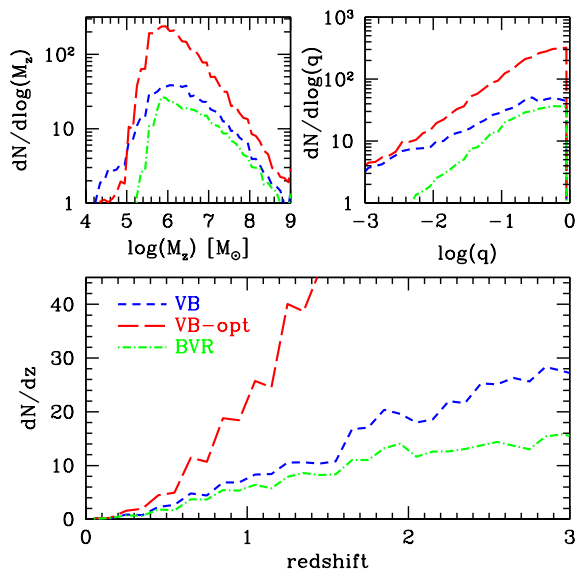}
\caption{Population of coalescing MBH binaries in three years. Top left panel: total redshifted 
mass distribution; top right panel: mass ratio distribution; lower panel: redshift distribution.
Color and linestyle codes are labeled in the figure.}
\label{mbhpop}
\end{figure}
The dashed blue lines in figure \ref{mbhpop} show the redshifted total mass ($M_z=(M_1+M_2)(1+z)$, being
$M_1>M_2$ the restframe masses of the two MBHs, upper--left panel),
mass ratio ($q=M_2/M_1$, upper--right panel) and redshift (lower panel) distribution of the MBH binaries
coalescing in three years, as seen from the Earth. 
The model predicts $\sim 40$ coalescences in the redshifted mass range $10^5\msun<M_z<10^7\msun$,
almost uniformly distributed in the mass ratio range $0.1<q<1$, with a long tail extending to $q<10^{-3}$.
For comparison we also show the population expected by a model featuring 
heavy seed only (BVR model, green dotted--dashed lines),
and by an alternative VB type model (labeled VB-opt for optimistic, red long--dashed lines) 
with a boosted efficiency of heavy seed formation \citep[see][for details]{Volonteri:2010py}.
It is worth mentioning that these models successfully
reproduce several properties of the observed Universe, such as the present day mass density of nuclear MBHs 
and the optical and X-ray luminosity functions of quasars \citep{mal07,sal07}. The BVR and the 
VB-opt models predict MBH population observables 
bracketing the current range of allowed values. The VB-opt model, in particular, is borderline  
with current observational constraints on the unresolved X-ray background, 
and it is shown here only for comparison.
In the following, we considered the VB model only, which fits 
all the relevant observables by standing on the conservative side. 

We performed 100 Monte Carlo realizations of the population of 
MBH binaries coalescing in three years. Each realization 
takes into account the distribution of the number of events and MBH masses with the redshift as predicted 
by the VB model. Other parameters (like time of coalescence, spins, initial orbital 
configuration) are chosen randomly using uniform priors over the appropriate allowed ranges.

\subsection{Galaxy distribution}
\label{SS:Sim:GAlDist}

To simulate the galaxy distribution in the Universe we use the data produced by the Virgo Consortium
publicly available at http://www.g-vo.org/Millennium.
These data are the result of the implementation of semi-analytic models for galaxy formation 
and evolution into the dark matter (DM) halo merger hierarchy generated by the Millennium simulation \citep{Springel:2005nw}. 
The Millennium run is a N-body simulation of the growth of DM structures in the expanding Universe starting 
from a Gaussian spectrum of initial perturbations in the DM field at high redshift, which successfully
reproduced the net-like structure currently observed in the local Universe.
The simulation has a side-length of $\approx 700$ Mpc (co-moving distance), and its
outcome is stored in 63 snapshots evenly separated in log($z$), enclosing
all the properties of the DM structure at that particular time. 
Semi-analytical models for galaxy formation are implemented {\it a posteriori} 
within the DM structures predicted by the simulation. 
Such models have been successful in reproducing several observed properties of
the local and the high redshift Universe \citep[see , e.g.,][]{bower06,De_Lucia:2006vua}. 
Here we use the implementation performed by Bertone and collaborators \citep{Bertone07}, 
which is a refinement of the original implementation by \cite{De_Lucia:2006vua}.

For each coalescing MBH binary, we choose the snapshot closest in redshift. 
Within the snapshot we choose the host of the GW signal according to a probability
proportional to the number density of neighbor galaxies $n_{\rm gal}$. Such assumption comes 
from the fact that two galaxies are needed in order to form a 
MBH binary, and we consider that the probability that a certain galaxy was involved in a galaxy merger is 
proportional to the number of neighbor galaxies. We consider to be neighbors of a specific galaxy 
all the $N(R)$ galaxies falling within a distance 
\begin{equation}
R =\sigma T_H(z),
\end{equation}
where $\sigma=500$ km s$^{-1}$ is the typical velocity dispersion of galaxies with respect to the 
expanding Hubble flow, and $T_H(z)$ is the Hubble time at the event redshift.
The number density of neighbor galaxies is then simply written as $n_{\rm gal}=3N(R)/(4\pi R^3)$.
When we choose the merger host, we compute $n_{\rm gal}$ considering {\it all} the neighbor galaxies,
without imposing any kind of mass or luminosity selection. In this case $n_{\rm gal} \equiv n_{\rm total}$.
However, when we will construct the probability of a given observable galaxy to be the host of 
the merger (i.e. the astrophysical prior $p_j(\theta, \phi, z)$), we will have to compute $n_{\rm gal}$ 
according to the number of {\it observed} neighbors,
because this is the only thing we can do in practice when we deal with an observed sample of galaxies
(see Section \ref{SS:Obs:zMeasure}).

\section{Simulating the observations}
\label{S:Obs}

\subsection{Gravitational wave observations: shaping the error box}
\label{SS:Obs:GWerrorbox}

As we mentioned in Section~\ref{SS:Sim:MBHBpop}, we drawn hundred realizations of the MBH binary population 
from the VB model. Each realization contains 30 to 50 events in the redshift range $[0:3]$. 
The total mass, mass ratio and redshift distributions of the events are shown in the figure~\ref{mbhpop}. 
In order to simulate GW observations, 
the binary sky location is randomly chosen according to a uniform distribution on the celestial sphere, 
the coalescence time is chosen randomly within the three years of {\it LISA} operation 
(we assume 3 years as default mission lifetime). 
the spin magnitudes are uniformly chosen in the interval $[0:1]$ in units of mass square, 
and the initial orientations of the spins and of the orbital angular momentum are  
chosen to be uniform on the sphere. More detailed description of the model for GW signal used in this paper is given in
\cite{2010PhRvD..81j4016P}.

The GW likelihood ${\mathcal L}$ needed in equation (\ref{Eq:like}) is approximated 
as a multivariate Gaussian distribution with inverse correlation
matrix given by the Fisher information matrix (FIM) :
\begin{equation}
\mathcal{L} \sim e^{-(s-h|s-h)} \sim e^{(\theta^i - \hat{\theta}^i) \Gamma_{ij}(\theta^j - \hat{\theta}^j)/2 }.
\end{equation}
Here $\theta^i$ is the vector of the parameters characterizing the spinning MBH binary, 
$\hat{\theta}^i$ are the maximum likelihood estimators for those parameters which are assumed to correspond
 to the true values (no bias), and $\Gamma_{ij} = (h_{,i}|h_{,j})$ is the FIM, where the commas correspond to
derivatives with respect to the parameters.
This is a reasonable approximation due to the large SNR
\citep[for more details on the FIM and its applicability see][]{Vallisneri2008}. 
Our uncertainties on estimated parameters are consistent with~\cite{Lang:2008gh}, ~\cite{2010CQGra..27h4009B} and~\cite{2010PhRvD..81j4016P}.
We did not include higher harmonics (only the dominant, twice the orbital frequency) as they only slightly
improve parameter estimation for precessing binaries. However including higher harmonics in the GW
signal model is important in case of the small spins and low precession (when spins are almost (anti)aligned 
with the orbital momentum, \cite{2011arXiv1101.3591L}).
We use truncated waveforms corresponding to the inspiral only. 
However the addition of merger and ring-down will further reduce the localization error 
 due to the higher SNR \citep{mcwilliams10}.
This error is usually an ellipse on the sky but we simplify it by 
choosing the circle with the same area. 

\begin{figure}[!htbp]
\center
\includegraphics[width=0.45\textwidth, clip=true]{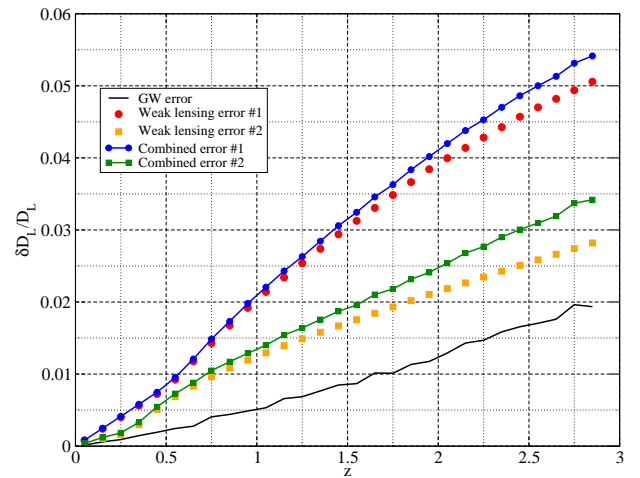}
\caption{Relative error in the luminosity distance due to weak lensing from (i) \cite{Shapiro:2009sr} (circles) and 
from (ii) \cite{Wang:2002qc} (squares). The black solid line is the median error due to GW measurements only; 
the solid-circle and the solid-square lines are for the combined errors under assumptions (i) and (ii) respectively 
(see text).}
\label{F:WL}
\end{figure}

For the luminosity distance measurement we need to take into account the weak lensing. 
We assume the weak lensing error to be Gaussian with a $\sigma$
given by (i) \cite{Shapiro:2009sr}. Such assumption is rather pessimistic; we also tried the prescription 
given by (ii) \cite{Wang:2002qc}, which gives smaller errors, but still larger than the level
that may be achieved after mitigation through shear and flexion maps \citep{Hilbert:2010am}. 
Both of those estimations are represented in figure~\ref{F:WL} as (i) dark (red online) 
circles and (ii) light (orange online) squares correspondingly.
The median error in $D_L$ due to GW measurements only is given by the solid black line. The combined error 
for model (i) is given by the upper (blue) circle-line curve, and for model (ii) by the lower (green) square-line 
curve.
We consider our setup to be conservative in the estimation of the weak lensing effects.
The main aim of this work is to build a reasonable setup for what could be observed by the time 
{\it LISA} will fly, and make a first order estimation of {\it LISA} capabilities to constrain 
the dark energy equation of state. We will address non-Gaussianity of the weak lensing as well as other 
corrections to the model to make it more realistic in a follow up paper.

We consider an error box size corresponding to $2\sigma$ of the measurement errors in the sky 
location ($\sigma_{\textrm{sky}}$) and in the source distance as evaluated by the FIM plus weak lensing uncertainties.
For observational purposes, the dimensions of this error box are $\Delta\Omega = 2 \sigma_{\textrm{sky}}$ and $\Delta{z}$.
For the latter we also include the uncertainty given in the $D_l-z$ conversion due to the error (prior) on $w$, $p_0(w)$.

\begin{figure}[!htbp]
\center
\includegraphics[width=0.5\textwidth, clip=true,clip=true]{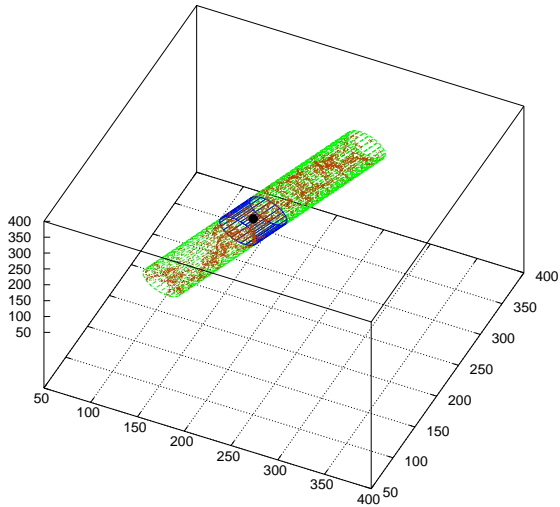}
\caption{Example of error box (cylinder) in part of the Millennium  snapshot (cube with unit in Mpc). The blue cylinder is the measurement error box and the green one also considers the prior on $w$. The black big dot is the host and the brown small dots are the selected galaxy candidates.}
\label{F:ExErrorBox}
\end{figure}

Let us summarize how we construct an error box in practice, as, for example, the one illustrated in figure~\ref{F:ExErrorBox}  :
\begin{itemize}
\item We select the closest Millennium snapshot to the event in redshift.
\item We pick a galaxy (red dot) in the snapshot with a probability given by the local galaxy number density $n_{\rm total}$.
\item We construct around the galaxy an error box given by $\Delta\Omega$ and $\Delta{z}_{GW+WL}$, and the galaxy can lie {\it anywhere} with respect to this error box (blue cylinder).
\item We expand the error box along the direction of the observer both sides by $\Delta{z}$ given by the uncertainty in $w$ (green cylinder).
\item According to some prescription,which we will describe in the next section, we select observable galaxies in the error box (brown dots).
\end{itemize}

As shown in figure \ref{F:ExErrorBox}, we interpret one of the directions in the Millennium snapshot as 
distance from the observer, and convert the comoving distance in redshift. 
We assume a periodic expansion of the Millennium data in order to fit large error boxes. Note 
that the original Millennium simulation also assumes the same periodicity in the distribution of the 
matter. The size of the error box at high redshift covers a significant fraction 
of the simulation box so we do not go beyond the redshift $z=3$ (as we will show later, 
spectroscopic observations at such high redshifts will be impractical anyway). Together with 
larger error boxes, we have a nonlinear increase in the number of events at high redshift. To reduce the 
overlap between error boxes corresponding to different GW events we choose cylinders with random orientations.

\begin{figure}[!htbp]
\center
\includegraphics[width=0.45\textwidth, clip=true]{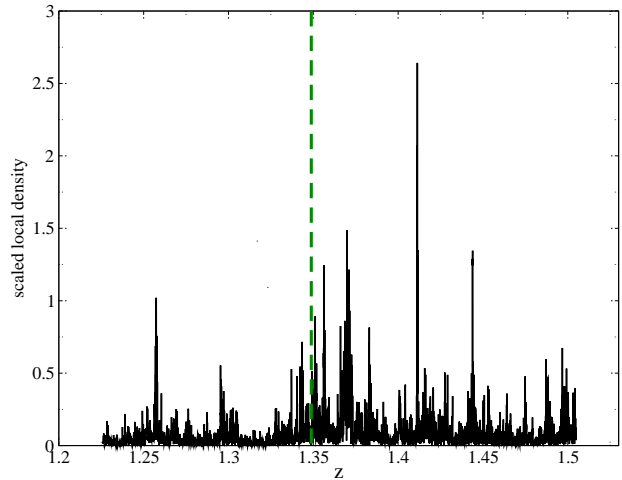}
\caption{Distribution of the weighted galaxies with the redshift. The green dashed vertical line is the redshift of the host galaxy.}
\label{F:ExZDistWGal}
\end{figure}

Figure~\ref{F:ExZDistWGal} shows an example of the resulting weighted distribution of galaxy redshifts
(with weight proportional to the local density $n_{\rm total}$).
It is a projection of the clumpiness along the line of sight which is also proportional to the probability distribution 
of $z$ for the event. The probability distribution of $w$ for the event will be directly related to this result.
We noticed that there is a very large number of underdense regions and several very dense superclusters. 
The probability of a galaxy with a low local density to host a merger is very low but there is a huge number 
of such galaxies, and we found that the probability of the host to be in (super)clusters is similar to that of being 
in a low density region. As we will see later in the result section, this may cause a very wrong estimation 
of $w$ for some individual GW event.

\subsection{Redshift measurements through spectroscopic surveys}
\label{SS:Obs:zMeasure}

To get a statistical measurement of $w$ we need to exploit the clustering
of the galaxies falling within the error box (which defines the astrophysical
prior $p_j(\theta, \phi, z)$ in equation (\ref{Eq:like})). It is therefore necessary
to get efficient redshift measurements of thousands of galaxies within 
a small field of view (FOV): the information we seek is enclosed
in the redshift distribution of such galaxies. We stress here that {\it we
are not looking for a distinctive electromagnetic counterpart to the GW event}. In fact,
the actual host of the coalescing binary may not even be observable.
Typical masses of our binaries are $10^5-10^6 M_{\odot}$. Using MBH-bulge
scaling relations \citep{gult09}, 
such MBHs are expected to be hosted in galaxies with
stellar mass $~10^{9}-10^{10}\msun$, i.e., in dark matter halos with total
mass $<10^{11}\msun$. The Millennium run mass resolution is 
$\sim10^9\msun$, meaning
that typical host structures are formed by less than 100 particles. Unfortunately,
the Millennium run is severely incomplete in the expected mass range of
{\it LISA} MBH binary hosts. Here we do not attempt to exploit any MBH-host
relation to select the host of our GW event; the probability of being a
host is only related to the local number density of neighbor galaxies
$n_{\rm total}$. Such 
assumption relies on the concept of {\it self-similarity} of the 
galaxy clustering at different mass scales: typical {\it LISA} MBH binary
hosts
cluster in the same way
as more massive galaxies. We checked this assumption by comparing the 
spatial distribution of galaxies in different mass ranges 
($10^{9}-10^{10}\msun$, $10^{10}-10^{11}\msun$, $10^{11}-10^{12}\msun$), within
simulation snapshots at different redshift, and we {\it postulate} that
this self-similarity extends to lower masses, below the Millennium run
resolution. This point is crucial
for two reasons: (i) especially at $z>1$, we will be able to get only 
spectra of luminous (massive) galaxies, and we need to be confident that 
their spatial distribution mimics that of lighter galaxies that may 
host the GW event but are observable in the spectroscopic 
survey; (ii) the number of observable galaxies in the error box may be
too large anyway ($>10^4$) to efficiently complete a spectroscopic survey
on the full sample: self-similarity allows us to get the clustering 
information we need by getting spectra of the brightest objects only. 

At $z=1$, the typical number of galaxies enclosed in the 2$\sigma$  error
box described above is in the range $10^4-10^5$. However, not all of them
are bright enough to get useful spectra. The semianalytic
galaxy evolution model \citep{Bertone07} implemented on top of the 
Millennium run returns the stellar mass of each galaxy, and the 
absolute bolometric magnitude $M_b$. By knowing the redshift, and by
using standard galactic templates one can therefore compute the apparent 
magnitude in a given band, by assuming 
the appropriate $k$ correction \citep{os68}. 
Here we use the $R$ band apparent magnitude $m_r$ for illustrative purposes, 
and we adopt the relation \citep{zombeck90}
\begin{equation}
M_b = -5{\rm log}(zc/H_0) - 1.086z - 25 + m_r + 0.6,
\end{equation}
where 0.6 is the $k$ correction. For each galaxy we compute $m_r$ 
and we simulate spectroscopic surveys at different thresholds $m_r=24, 25, 26$. 
We stress here that the GW host was chosen
among {\it all} the galaxies falling in the error box, and therefore may 
not (and usually does not) belong to the observed sample. We then 
assume that for each galaxy satisfying the survey threshold we get an 
exact spectroscopic redshift, and we combine the redshift distribution 
of several error boxes to get a statistical estimation of $w$. In practice,
each redshift estimation will come with a measurement error, and an intrinsic
error due to the proper motion of the source with respect to the Hubble flow. 
Both errors are however of the order of $\Delta{z}/z<10^{-3}$, well below the 
typical redshift scale corresponding to spatial clustering of structures 
($\Delta z\sim0.01$, see figure \ref{F:ExZDistWGal}) we need to resolve.

Our method does not rely on the observation of a prompt 
transient associated to the MBH binary coalescence to identify the 
host galaxy. Nevertheless, getting thousands (or tens of thousands) 
of spectra in a small field of view requires a dedicated observational
program. Thanks to multi-slit spectrographs such as VIMOS at VLT \citep{lefevre03}
and DEIMOS at Keck \citep{faber03}, fast deep spectroscopic 
surveys of relatively large FOV are now possible. 
For example, the ongoing VIMOS VLT deep survey \citep{lefevre05}, 
took spectra of $>10000$ galaxies, mostly in the redshift range $0<z<1.5$, 
within a FOV of 0.61deg$^2$ at an apparent magnitude limit 
$I_{AB}<24$. Comparable figures are achieved by other 
observational campaigns such as zCOSMOS \citep{lilly09} 
and DEEP2 \citep{davis03}, being able of surveying selected galaxies 
in various photometric bands ($U, B, R, I$) to an apparent magnitude 
limit of about 24. Going deeper in redshift, Lyman break galaxy redshift 
surveys are being
successful in efficiently getting high quality spectra of hundreds of galaxies
in the redshift range $2.5<z<3.5$ within a FOV $\sim1$deg$^2$ \citep{bielby10}. 
To get an idea, the VIMOS spectrograph 
can take $\sim 500$ high quality spectra per pointing with an integration
time of about $4$h, within a 7$\times$8 arcmin$^2$ FOV, which is 
coincidentally of the same order of the typical error box for a $z=1$
GW event. The typical redshift accuracy of the spectra
is $\Delta{z}<10^{-3}$ ($3\times10^{-4}$ in the zCOSMOS survey, $2\times10^{-3}$
in the Lyman break galaxy survey), well below the typical redshift scale 
we are interested in ($z\sim0.01$). 

Such figures witness the feasibility of efficient
spectroscopic redshift determination of a large sample of galaxies at
faint apparent magnitude ($m_r \approx 24$), as required by our problem. 
Future spectroscopic survey as BigBOSS \citep{bigboss09} are expected 
to further improve such figures of merit; a new spectrograph will be able
to simultaneously get up to 4000 spectra within a single pointing of a 
7deg$^2$ FOV. Getting few thousand spectra of objects falling within 
the GW error box in the redshift range of interest may be possible 
in a single observing night. At a $m_r=24$ cut-off magnitude 
we generally have few hundred to few thousands galaxies in the GW 
error box, but we go deeper (i.e., $m_r=26$, feasible with future
surveys), the number of spectra may increase drastically. For some
of the error boxes, we count up to $10^5$ galaxies with $m_r<26$.
However, the requirement of a factor of ten more spectra, does not correspond to a
significant improvement of the results. This is a consequence of the self similarity 
of the galaxy distribution: as long as there are enough galaxies in the error box 
to recover the clustering information, the results are basically independent on the 
assumed cut-off magnitude. A survey with a cut-off magnitude of $m_r=24$ may indeed 
be a good compromise between reliability of the results and time optimization in terms 
of follow-up spectroscopy.

The magnitude cut-off defines the number of neighbor observable galaxies.
This is the only practical way to weight each galaxy with a local density, 
$n_{\rm gal} \equiv n_{m_r}$ (the subscript $m_r$ refers to the adopted magnitude limit) 
along the lines discussed in Section \label{SS:Sim:GAlDist}. 
Once we have a spectroscopic galaxy sample, each galaxy in the error box comes with the prior probability 
to be the host proportional to $n_{m_r}$, so the astrophysical prior in equation
(\ref{Eq:poster}) could be written as 
\begin{equation}
p(\Omega, z) = \sum_i n_{m_r,i} \; \delta(\Omega - \Omega_i) \delta(z - z_i)
\end{equation}
where the sum is over all observable galaxies in the error box and $\Omega$ 
is the geodesic distance on the celestial sphere from the center of the box. 
At redshifts $z \ge 1$  the prior probability $p(\Omega, z) $ becomes almost a 
continuous function (as the example in figure~\ref{F:ExZDistWGal}). 


\subsection{Approximations and caveats}
Before jumping to the results, we want to mention some corrections we made to accommodate the limitations of our simulations.
Firstly, we interpreted one of the directions in the snapshot (along the side of the cylinder) as distance from the 
observer. This is a good approximation only if the error box size is small.  
For large error boxes, a uniform distribution in the 
comoving distances does not translate into a uniform distribution in redshifts: there is an artificial slope with a bias 
toward low values of $z$. We have corrected for this slope. Secondly,
the clumpiness evolves with redshift, which is not the case if we use a single snapshot and interpret 
one of the directions as a redshift. To properly account for this, we should glue snapshots together 
and perform an interpolation between them.  However we wanted to simplify the setup for 
this very first attempt. The main idea was to check whether the 
density contrast within the error boxes is sufficient to constrain further the error on $w$. If the distribution of density 
within the error box is uniform then we do not gain any useful information. However there is a natural bias: for a given 
measurement of $D_L$, the galaxy further away (larger $z$) constrains $w$ better than galaxy at lower redshift. 
One can see it from the fact that deviation between the curves in $D_L - z$ plane corresponding to the small 
deviation in $w$ is bigger for large $z$. This could be counterbalanced by the decreasing density contrast at large redshift. 
Here, we corrected the slope of the posterior $P_j(w|s)$ by demanding that a uniform distribution 
$p_j(\theta, \phi, z)$  returns a posterior on $w$ equal to the prior, i.e., $P_j(w|s) = p_0(w)$.

\section{Constraints on the dark energy equation of state}
\label{S:Res}

In this section we present the results of our simulations. We tried several 
setup of the experiment by using different thresholds on the observable apparent magnitude of 
galaxies, different prescriptions for the measurement errors, and different cosmological priors.
For each setup, we performed either 100 or 20 realizations of the MBH binary population
as observed by {\it LISA}, together with the follow up spectroscopic survey of the galaxies in all the
error boxes.
 
\subsection{Fiducial case}
\label{SS:Res:General}
We consider in this subsection 100 realizations which we refer to as our fiducial case. 
For this setup, we limit 
spectroscopic identification of galaxies in the error box to an apparent magnitude of 
$m_r \le 24$, the errors in sky localization and in the luminosity distance are estimated according 
to the inspiral part of GW signal only, and the weak lensing uncertainty is taken from \cite{Shapiro:2009sr}.
The prior $p_0(w)$ was assumed to be uniform $U[-0.3:0.3]$ with  an exponential decay at the boundaries. 
Such interval is consistent with current $2\sigma$ (95\% confidence level) constraints on $w$ 
\citep[$w=-0.12\pm0.27$,][]{2010arXiv1001.4538K}, obtained by 
cross correlating seven-year WMAP data with priors coming from independent measurements of $H_0$ and
barionic acoustic oscillations \citep[see][and references therein for full details]{2010arXiv1001.4538K},
under our same assumption for the dark energy equation of state, $\omega = -1 - w$, where $w$
is a constant. Such range is reduced by a factor of almost three ($w=-0.02\pm0.1$) when type Ia supernovae 
data \citep{riess98} are included. Here we show that GW measurements offer a competitive alternative to
type Ia supernovae, placing an independent constraint on the dark energy equation of state.

\begin{figure}[!htbp]
\center
\includegraphics[width=0.45\textwidth, clip=true]{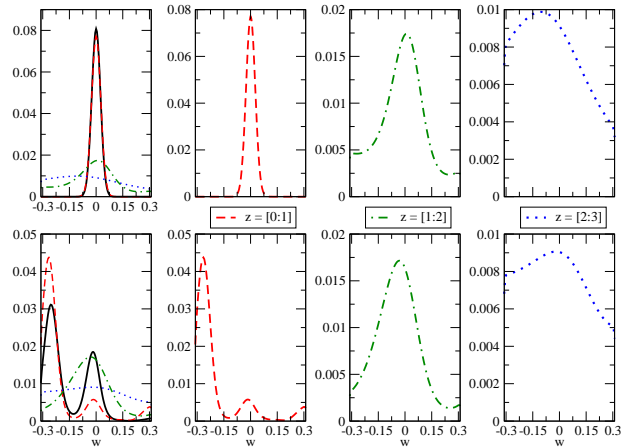}
\caption{Posterior distribution for $w$ for two particular realizations (top and bottom row). In each row, 
the left plot shows the full posterior from 
all GW events (black curve) as well as contributions from different redshift bands. The three right plots show the individual contribution for the three redshift ranges, as labelled in the panels. 
}
\label{F:wPostFiducial2Ex}
\end{figure}

We find that in almost all cases we improve the constraints on $w$,
in other words, the posterior distribution is narrower than the prior. 
Few events at low redshift usually play a major role in the final result. 
One typical realization is plotted in the top panel figure~\ref{F:wPostFiducial2Ex}.
We split the contribution to the posterior distribution $P(w)$ in
redshift bands: $z\in [0:1]$ (second plot from the left), $[1:2]$ (third plot), $[2:3]$ (fourth plot).
Their relative contribution and the resulting posterior (black) is given in the leftmost plot.
In this example the final posterior probability is almost completely determined by few events at low redshift.
The second realization, shown in the lower panels of figure~\ref{F:wPostFiducial2Ex}, demonstrates how low redshift 
contributions could give inconclusive results. In this particular case, there are two maxima with preference 
given to the wrong one. The contribution from high redshift events could change this ratio as it is 
shown in this example. In many cases the mergers above redshift $z=1$
can constrain $w$ only to a 0.1-0.15 accuracy, but they almost always add up coherently giving a 
maximum at the right value ($w=0$). 
This usually helps in case the low redshift events return a multimodal $P(w)$, and is, in turn,
the power of our statistical method.  

\begin{figure*}[!htbp]
\begin{tabular}{ccc}
\includegraphics[scale=1.03,clip=true,viewport=13 2 167 110]{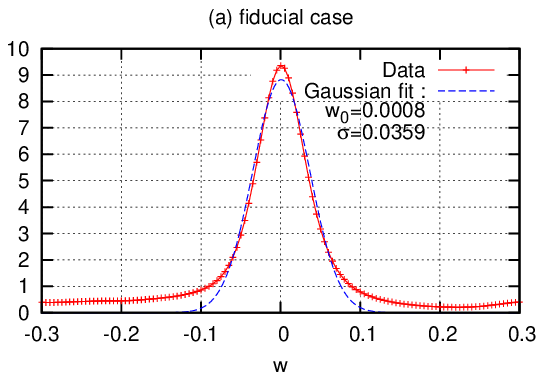}&
\includegraphics[scale=1.03,clip=true,viewport=13 2 167 110]{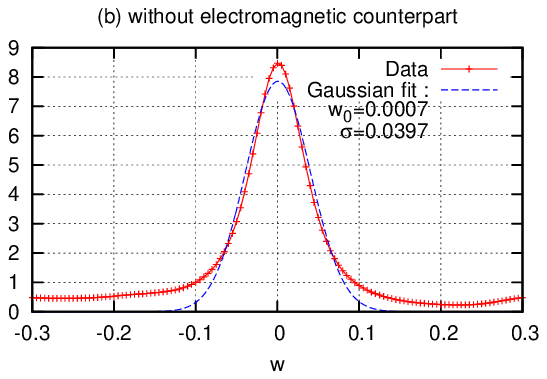}&
\includegraphics[scale=1.03,clip=true,viewport=13 2 167 110]{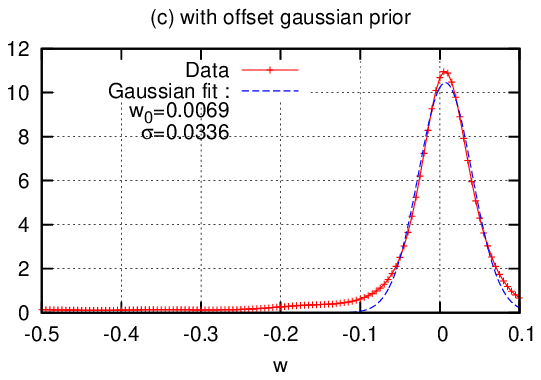}\\
\includegraphics[scale=1.03,clip=true,viewport=13 2 167 110]{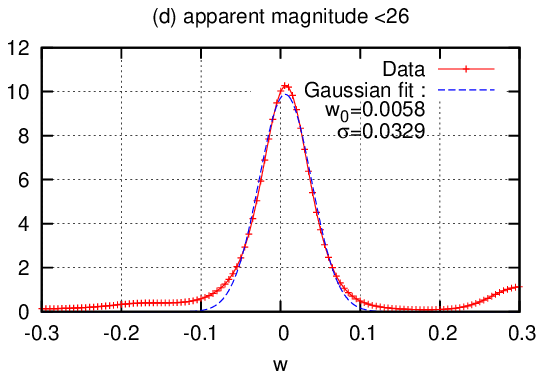}&
\includegraphics[scale=1.03,clip=true,viewport=13 2 167 110]{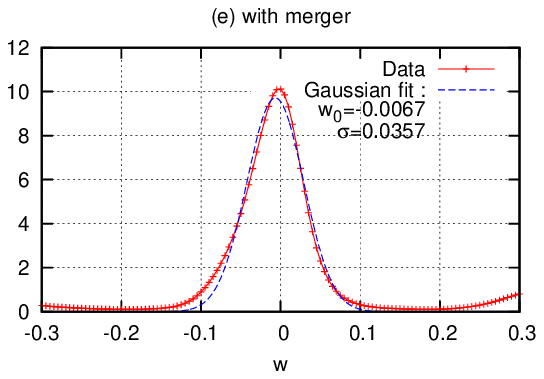}&
\includegraphics[scale=1.03,clip=true,viewport=13 2 167 110]{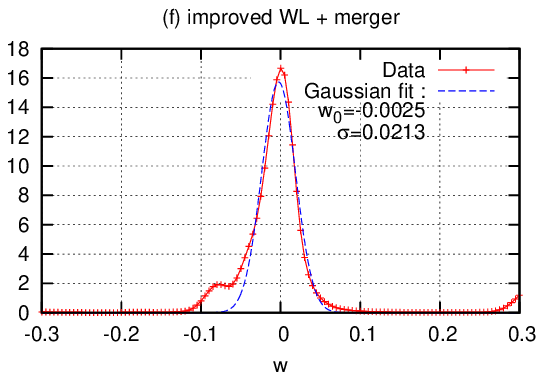}
\end{tabular}
\caption{Collective figures of merit of our experiment. 
In each panel, corresponding to a different setup of our experiment as labelled in figure, 
the red solid curve corresponds to the data, i.e. the sum of the posterior distributions of $w$ over 
all realizations. The blue dashed curve is a Gaussian fit with parameter given in the legend of each plot.
}
\label{F:ResFigOfMerit}
\end{figure*}

\begin{figure}[!htbp]
\center
\includegraphics[width=0.45\textwidth, clip=true]{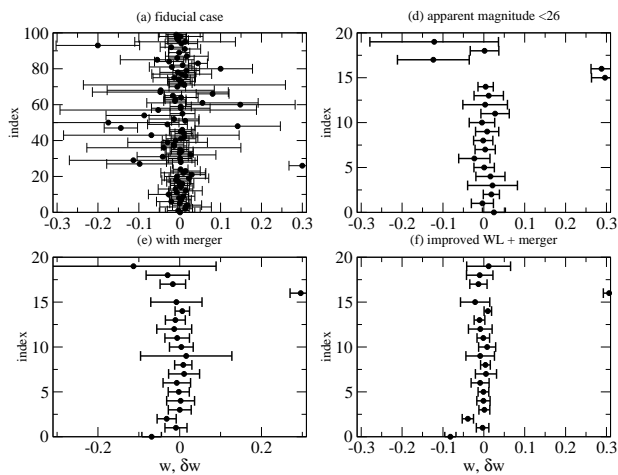}
\caption{Mean values and standard deviations resulting from the Gaussian fit 
of the posterior $P(w)$. The setup of each panel correspond to the one adopted in the panel of 
figure ~\ref{F:ResFigOfMerit} labelled by the same letter).}
\label{F:ResMeanVar}
\end{figure}

We characterize the results of each setup (100 or 20 realizations) using the figures of merit 
shown in figures~\ref{F:ResFigOfMerit} and \ref{F:ResMeanVar}.
The first one (figure~\ref{F:ResFigOfMerit}) is obtained by adding the posterior distributions $P(w)$ of  
all the realizations.
We fit the resulting curve with a Gaussian, characterizing the result using its
mean $w_{0}$ and standard deviation $\sigma_{w}$. The second figure of merit (figure~\ref{F:ResMeanVar})
shows the result of Gaussian fits performed on each individual realization (vertical index $i$): 
the mean $w_{0}(i)$ is shown as a circle and the standard deviation $\sigma_{w}(i)$ is the error bar. 
The first figure of merit gives collective information, showing how well, on average, an individual 
realization can be approximated by a Gaussian fit, while the second figure of merit shows the dispersion 
of the posterior distribution across the individual realizations. 

The fiducial case, featuring 100 realizations, is shown in panel (a) of both figures
~\ref{F:ResFigOfMerit} and \ref{F:ResMeanVar}. The parameters of the global fitting Gaussian mean are 
$w_{0} = 0.0008$ and $\sigma_{w} = 0.036$, corresponding to a factor of four improvement in 
the estimation of $w$ with respect to our standard $2\sigma$ $[-0.3:0.3]$ prior. 
However the distribution has clearly some outliers, recognizable as non-Gaussian 
tails in figure~\ref{F:ResFigOfMerit} and pinned down in figure~\ref{F:ResMeanVar}.
For the fiducial case,  
84\% of the realizations have a mean value close to the true one, i.e. $ | w_{0}(i) - w_{\rm true} | < 0.1$ 
with an appreciable reduction of the prior range, i.e. $\sigma_{w}(i) < 0.15$ ($i=1,..,100$ 
is the realization index). Moreover, most of the outliers can be corrected as we will 
explain in Section~\ref{SS:Res:CheckQual}. 

%

\subsection{Removing ``electromagnetic counterparts''}
\label{SS:Res:NoEMC}
Our goal is to demonstrate that we are able to constrain the dark energy equation of state 
{\it without} directly observing electromagnetic counterparts. However, for some of the low redshift 
events, the error box is so small that only one or two galaxies fall within it. 
Having one or two galaxies in the error box essentially implies an electromagnetic identification of the host,
so we decided to re-analyze the fiducial case removing
all such fortunate events (usually 0-2 in each realization). The fiducial case without clearly
identifiable hosts is presented in the panel (b) of figure ~\ref{F:ResFigOfMerit}.  
Clearly, our results remain almost unchanged, the posterior distribution is slightly wider (larger sigma) and
non-Gaussianity is more pronounced.

\subsection{Choice of the prior for $w$}
\label{SS:Res:wPriors}
Here and in the next subsections we make use of 20 selected realizations, which we found to be 
sufficient to depict the relevant trends of the analysis. We took 15 ``good'' (mean values close to the true 
and small rms errors) and 5 ``bad'' cases from the fiducial setup.

In this subsection we study the effect of the prior $p_0(w)$ on the posterior distribution. 
We considered an extreme case: a Gaussian $\mathcal{N}(w_{0}=-0.2, \sigma=0.3)$.
As shown in panel (c) of figure~\ref{F:ResFigOfMerit},
the global posterior distribution is still centered at the true value $w=0$.
This demonstrates that the final conclusion is basically unaffected by the choice of the 
prior (as long as the prior covers the true value) and GW observations, in principle, could be used as 
an independent mean of estimating $w$.

\subsection{Using deeper surveys}
\label{SS:Res:Mag26}
Here we study the dependence of our results on the depth of the follow up spectroscopic survey:
i.e. on the observability threshold. We considered the same 20 realizations as in the previous section,
but now with different limits on the apparent magnitude of observable galaxies: $m_r = 24, 25, 26$.
The case $m_r = 26$ is given in panel (d) of figures~\ref{F:ResFigOfMerit} and~\ref{F:ResMeanVar}.
The results are comparable to the fiducial case. They show a small improvement 
in sigma and slightly larger bias for the combined distribution. We also notice
that 4 out of 5 ``bad'' cases remain bad.

We should say few words about the number of galaxies used here.
As mentioned above, the typical number of galaxies for the fiducial case 
($m_r = 24$) is less than few thousand  for events at $z<1$
and less than few tens of thousands for the high redshift event.
For the improved observational limit ($m_r = 26$), these numbers are 2 to 10 times larger.
The fact that our results are not sensitive to the depth of the survey reflects 
the self-similarity of the spatial distribution of galaxies in different mass ranges.  


\subsection{Improving the sky localization and the luminosity distance estimation}
\label{SS:Res:Improve}
In our fiducial setup, the assumed source sky localization and luminosity distance error
are rather conservative. In this subsection we consider the effect of improving such measurements. 
So far, we considered only the inspiral part of the GW signal; the inclusion of merger and ringdown 
will improve the localization of the source by at least a factor of two \citep{mcwilliams10},
due to the large gain in SNR. We artificially reduced the sky localization error coming 
from the inspiral by a factor of two (factor of four in the area), assuming that this will be the case if we 
take the full GW signal. 
We reanalyzed the same 20 realizations with this new error on the sky.
Because the size of the error box is smaller, the number of potential counterparts is reduced by a factor of $\sim4$ compared to the fiducial case.
The results are presented in panel (e) of figure~\ref{F:ResFigOfMerit}. 
We see that the main effect of a better GW source localization 
 is to reduce the number of outliers and to remove the non-Gaussian tails in the combined probability.
As it is clear form panel (e) of figure~\ref{F:ResMeanVar}, the main gain comes from improvement
of the ``bad'' cases.


We now consider another estimation of the mean weak lensing contribution to the 
luminosity distance error, given in~\cite{Wang:2002qc} 
(green square-line curve on figure~\ref{F:WL}). We take this in combination with improved source 
localization on the sky coming from taking into account the merger (as discussed above).
We consider the same 20 realizations. Results are shown in panel (f) of both
figures ~\ref{F:ResFigOfMerit} and ~\ref{F:ResMeanVar}. The improvement with respect to all
the other cases is obvious.
Because the marginalized likelihood  $\pi_{j}$ coming with each galaxy is narrower 
due to the smaller error in the luminosity distance, 
the final posterior on $P_{j}(w)$ is also narrower. 
The standard deviation $\sigma_{w}$ is improved by more than 40\% as compared to the fiducial case. 
The non-Gaussian tails have almost completely disappeared, due to the removal of the outliers 
(further improvement of the ``bad'' cases, the remaining bad case will be treated in the subsection \ref{SS:Res:CheckQual},
see also the top panel of figure \ref{F:SelfConsistency}). With this model of the mean weak lensing contribution  
and assuming the full GW signal, the estimation of $w$ is improved by a factor of $\sim8$ 
as compared to the initial uniform prior.

\subsection{Consistency check.}
\label{SS:Res:CheckQual}

\begin{figure}[!htbp]
\center
\includegraphics[width=0.45\textwidth, clip=true]{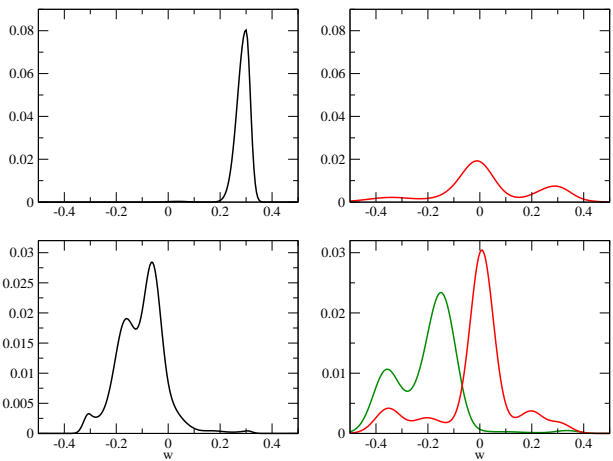}
\caption{Each row of panels show our self-similarity check for a selected realization. 
In each row, the solid curve on the left panel corresponds the final posterior $P(w)$
The solid curves on the right panel
are the posteriors after removing one event, $\widetilde{P}_{k}(w)$.
}
\label{F:SelfConsistency}
\end{figure}

As we mentioned above, some nearby GW event could seriously bias the final posterior. 
We also mentioned that the odds for the 
host to  be in a low density region of the Universe are not small. The posterior
probability $P(w)$ reflects the distribution of the mass defined by the
astrophysical prior $p_j(\theta, \phi, z)$. A nearby GW event hosted in the 
low density environment could seriously damage the final result. An example 
is given in the top left panel of figure~\ref{F:SelfConsistency}.
In order to eliminate or at least test such unfortunate cases we performed a 
self-consistency test on our results. Basically we remove one GW event from the 
analysis and see if the resulting posterior $\widetilde{P}_{k}(w)$ distributions are consistent. 
We defined the posterior of all the events minus one as:
\begin{equation}
\widetilde{P}_{k}(w) = \frac{p_0(w) \prod_{j \neq k} P_j(s| w)} {\int p_0(w) \prod_{j \neq k} P_j(s | w) dw}. 
\label{E:MultiplyPost}
\end{equation}
If  $\widetilde{P}_{k}(w)$ give similar results for all $k$, then we can be confident that the result is not 
biased by one particular unfortunate event, and this increases our trust in the final posterior distribution.
If, conversely, all $\widetilde{P}_{k}(w)$ but one are consistent, then we say that this one event 
is not in line with the remaining events and should be abandoned. In the top panels of  
figure \ref{F:SelfConsistency} we see
that removing one event at low redshift changes the final probability completely; 
the solid (red) line in the right panel is the new posterior distribution, consistent 
with the true value $w=0$.
However, there are still few cases where the self-consistency test is not conclusive, 
and one of them is shown in the lower panels of figure \ref{F:SelfConsistency}.
In this case, removing one ``bad'' nearby event produces the red curve centered at $w=0$, but removing another 
(``good'') event results in the green curve, which are mutually not consistent at all. Since in real life we will
not know which event is ``good'' and which one is ``bad'', we will not be able to make a clear definite statement, 
and our answer will be bi-modal with a probability attached to each mode.

\subsection{Comparison with the optimal case: detection of electromagnetic counterparts.}
\label{SS:Res:DetectEM}

For comparison, we have also considered the best possible case, in which the redshifts of the GW 
source hosts are determined unambiguously through the identification of a distinctive electromagnetic
counterpart. In this case, the redshift of each GW event is known exactly (within negligible measurement
errors). Therefore, the error on $w$ comes only from the error on luminosity distance 
(GW error measurement plus weak lensing).
Considering 20 realizations with a configuration equivalent to the fiducial case
(Section~\ref{SS:Res:General}), the global posterior distribution is a Gaussian centered 
at $w_{0} = 0$ with $\sigma_{w} =0.021$ (for comparison, see panel (a) of figure \ref{F:ResFigOfMerit}). 
With a configuration equivalent to our improved case, i.e. better weak lensing
(Section~\ref{SS:Res:Improve}),
we obtain $\sigma_{w} = 0.012$ (for comparison, see to panel (f) of figure \ref{F:ResFigOfMerit}). 
In both case the difference between our statistical method and the 
best possible case (all electromagnetic counterparts detected) is only about a factor 2.



\section{Summary}
\label{S:Sum}

In this paper, we presented a statistical method for constraining 
cosmological parameters using {\it LISA} observations of spinning massive black hole binaries
and redshift surveys of galaxies.
Our approach does not require any direct electromagnetic counterpart; instead, the
consistency between  few dozen of GW events imposes constraints on the redshift-luminosity distance 
relationship. This, in turn, allows us to estimate cosmological parameters. This method strongly relies 
on the non-uniformity (i.e., clustering) of the galaxy distribution within the uncertainty error box set by {\it LISA} 
observations, weak lensing and priors on the cosmological parameters.

For this first exploratory study, we fixed all the cosmological parameters but one, 
$w$, describing the effective equation of state 
for the dark energy. We used the Millennium simulation to model the Universe at different redshifts. 
We used a particular (VB) hierarchical MBH formation model to mimic the MBH binary population observed
by {\it LISA}. Using this setup, we considered between 20 and 100 realizations of 
the observed {\it LISA} binary population. We tried two different models for estimating the error in 
luminosity distance due to weak lensing, we also looked at the effect of including merger and ringdown via improvement 
of the sky localization. We checked the robustness of our final result against different
depth of future spectroscopic galaxy surveys. 

Our fiducial case, based on conservative assumptions, shows that we are able to constrain 
$w$ to a 8\% level ($2\sigma$), i.e., we improve its estimate by a factor of $\sim4$ as 
compared to the current 95\% confidence interval obtained by cross correlating the seven-year WMAP data analysis with
priors coming from $H_0$ measurements and barionic acoustic oscillations \citep{2010arXiv1001.4538K}.
Such new measurement would be at the same level (25\% better on average) than current constraints based
on seven-year WMAP data plus type Ia supernovae observations. The optimistic case (smaller weak lensing disturbance 
and full GW waveform) allows us a further improvement by another factor of two, providing a 
factor of $\sim2.5$ tighter constraint than current estimates including supernovae data. 
Our results are most sensitive to the weak lensing  error (witnessing once more how critical is the issue of 
weak lensing mitigation for cosmological parameter estimation through GW observations) and are almost 
independent on the depth of the redshift survey (provided we have a reasonable number of redshift measurements per error box). 

In the majority of the realizations the most information comes from few events at low redshift, 
and high redshift events do help in case of multimodal structures in the posterior distribution. 
We suggested a self-consistency check based on the similarity of the posterior distribution from each 
GW  event. This increases our confidence in the final result and allows to reduce the risk 
of incurring in unfortunate outlier realizations for which we can not place useful constraints
on $w$. We also compared our statistical method to the optimal situation in which electromagnetic 
counterparts to the GW sources are identified, finding an improvement of a factor of two in the 
latter case. In absence of distinctive electromagnetic counterparts, statistical methods like the one 
presented here can still efficiently constrain cosmological parameters.

Although the main result of the present paper is encouraging, it was obtained assuming a fixed cosmological model
with one free parameter only: the $w$ parameter describing the dark energy equation of state. Even though we will likely have a good knowledge of most of the other cosmological parameters by the time {\it LISA} will fly, 
it is worth considering models with more degrees of freedom. In following studies, we intend to consider a more 
realistic situation by releasing other cosmological parameters, testing {\it LISA} capabilities of setting constraints 
on a multi parameter model.

\begin{acknowledgments}
Work of A.P. and S.B. was supported in parts by DFG grant SFB/TR 7 Gravitational Wave Astronomy and by DLR
(Deutsches Zentrum fur Luft- und Raumfahrt). The Monte-Carlo simulations were performed on the Morgane cluster
at AEI-Golm and on the Atlas cluster at AEI-Hannover. The authors would like to thank Jonathan Gair and Toshifumi Futamase 
for useful discussions.
\end{acknowledgments}

\bibliographystyle{apj}

\bibliography{refr}

\end{document}